\documentclass[preprint,aps,10pt,prb,twocolumn,showpacs,floatfix]{revtex4}
\usepackage{graphicx} 
\usepackage{epsfig} 
\begin{document}

\title{Computer simulation of the phase diagram for a fluid confined  
  in a fractal and disordered porous material}

\author{
V.~De Grandis$^\dagger$, P.~Gallo$^\dagger$ and
M.~Rovere$^\dagger$\footnote[1]
{Author to whom correspondence  should be addressed; e-mail: 
rovere@fis.uniroma3.it}}
\affiliation{$\dagger$ Dipartimento di Fisica, 
Universit\`a ``Roma Tre'', \\ 
INFM and Democritos National Simulation Center,\\ 
Via della Vasca Navale 84, 00146 Roma, Italy.}

\begin{abstract}
We present a grand canonical Monte Carlo simulation study of the
phase diagram of a Lennard-Jones fluid adsorbed in a fractal 
and highly porous aerogel. The gel environment
is generated from an off-lattice diffusion
limited cluster-cluster aggregation process.
Simulations have been performed with the multicanonical 
ensemble sampling technique. The biased sampling function
has been obtained by histogram reweighting calculations. 
Comparing the confined and the bulk system liquid-vapor coexistence 
curves we observe a decrease of both the critical temperature and 
density in qualitative 
agreement with experiments and other Monte Carlo studies 
on Lennard-Jones fluids confined in random matrices of spheres.
At variance with these numerical studies we do not observe upon confinement  
a peak on the liquid side of the coexistence curve 
associated with a liquid-liquid phase coexistence. In our case
only a shouldering of the coexistence curve appears 
upon confinement.  
This shoulder can be associated with high density fluctuations 
in the liquid phase. The coexisting vapor and liquid phases in our system 
show a high degree of spatial disorder and inhomogeneity.
\end{abstract}


\maketitle

\section{INTRODUCTION}

The phase behavior of fluids and fluid mixtures confined 
in porous and disordered materials represents a field
of continuing theoretical and experimental interest
due to a variety of applications in industrial technology~\cite{gelb}.
For example, porous materials are employed as adsorbents 
in many industrial processes, such as catalysis,
adsorption separation, filtration and purification.

Experimental studies on the phase behavior of fluids adsorbed
both in highly porous materials, such 
as silica aerogels~\cite{maher,wong1,frisken1,frisken2,wong2,frisken3}, 
and in less porous ones, such as Vycor 
glasses~\cite{wiltzius1,finotello,wiltzius2,wiltzius3,evans,lin} ,
have been performed.
All these studies have shown that the phase diagram of the fluids confined
in the porous structures is strongly 
altered in comparison with the corresponding bulk systems.
For $^4He$~\cite{wong1} and $N_{2}$~\cite{wong2} confined in very 
dilute aerogels it has been found that the liquid-vapor coexistence 
curve is much narrower and the critical temperature is lower than in the bulk.
It has also been observed a shift of the critical density 
towards the liquid phase and an increase of the coexistence
vapor phase densities attributed to the attractive fluid-gel interactions.

Very few theoretical studies on fluids confined in disordered 
porous materials have been performed. 
A theoretical approach based on considering 
the gel as a random field acting on the fluid
succeeded in reproducing the important feature of spatial 
inhomogeneity~\cite{degennes}. Important phenomena like wetting
cannot nonetheless be reproduced by such model.
A model of ``quenched-annealed'' binary mixture 
was first studied by Madden and Glandt~\cite{madden} and then 
widely used in dealing with the problem of fluids phase separation 
in porous materials.
Following this model, several integral equation theories and computer
simulation studies have been successively
proposed, in which the porous material is
described by a random matrix of 
spheres~\cite{rosinberg1,rosinberg2,toigo,monson1,alvarez}.

In particular the simulation work done by Page and Monson~\cite{monson1}
dealt with the calculation of the phase diagram for a system
representative of methane adsorbed in a silica xerogel 
and that of Alvarez and al.~\cite{alvarez}
has investigated the sensitivity 
of the confined fluid phase behavior to the matrix realization.
The results of these studies about the vapor-liquid coexistence curve 
are in good qualitative agreement with experiments. 
The novelty of these works not observed in experiments 
is the occurrence of an additional phase transition
between a medium density liquid and an high density one.
The liquid-liquid phase coexistence has been associated
with the wetting properties of the fluid in the more dense 
regions of the adsorbent. 
This second phase transition however has been found to 
be very sensitive to the 
matrix realization, while the liquid-vapor coexistence properties
are seen to be robust to variations 
in the solid structure~\cite{alvarez,monson2}.
This behavior has been explained observing that the additional 
liquid-liquid transition is associated 
with the filling of low porosity regions in the matrix, which
can occur for some configurations but not for others.
The picture which emerges from all these calculations is that 
the coexistence vapor and liquid phases for the confined fluids are 
disordered and inhomogeneous as a consequence of the
adsorbent structure randomness.

The description of the aerogel as a random matrix of spheres however
is not sufficiently realistic. The gel is organized into a fractally 
correlated structure experimentally identified by a strong 
diffraction peak at small wave vectors measured by small angle x-ray or
neutron scattering~\cite{keefer,freltoft,hasmy1,hasmy2}.
The fractal behavior of aerogels is associated with the
irreversibility of the gel formation dynamics, 
which proceed by random colloidal
aggregation of silica particles. The fractal dimension of silica
aerogels is about 1.80 in three dimensions~\cite{hasmy1}.

Structures with similar fractal dimension have been generated
by Monte Carlo simulations using several hierarchical cluster-cluster
algorithms~\cite{hasmy1,hasmy2,hasmy3}.
In particular the  Diffusion Limited Cluster-Cluster 
Aggregation procedure (DLCA) is a gel growth process  
widely used in theoretical studies of aerogels~\cite{kolb,hasmy2}. 
It has been shown that the structural properties 
and fractal dimension of DLCA gels well 
agree with experimental data on silica aerogels.
A theoretical study of thermodynamical and structural properties of a 
Lennard-Jones fluid adsorbed in a highly porous
DLCA aerogel by means of an
integral equation approach has been recently proposed~\cite{rosinberg2}.
The phase diagram as well as the structural correlations of the confined
fluid are found to be influenced by the specific properties of the gel,
such as its connectivity and fractal behavior, particularly
at low fluid densities.
These results support the idea that a realistic modeling of the
gel environment is necessary to deeply investigate the effects of the porous 
medium structural properties on the phase behavior of the adsorbate, 
especially on the second disorder-induced liquid-liquid transition.

In this paper we present the results of a grand canonical 
Monte Carlo simulation study about the 
phase behavior of a Lennard-Jones fluid confined in a dilute
DLCA aerogel, where fluid and gel particles have hard core diameters
of equal size and interact by means of an hard sphere potential.
In order to locate the liquid-vapor coexistence curve
we have calculated the density distribution functions.
The multicanonical ensemble sampling (MES) procedure
has been employed to investigate thermodynamical 
states in the subcritical region~\cite{berg,wilding1,wilding2}. 
In the next section we describe the algorithm to build our confining system.
In Sec.~3 we give details of our simulation.
In Sec.~4 we present the results about the confined fluid 
phase diagram and compare it with that of the bulk system. 
Sec.~5 is devoted to conclusions.
 
\section{CONFINING SYSTEM}

The algorithm we used to generate the aerogel configuration
is the three dimensional 
off-lattice extension by Hasmy and al.~\cite{hasmy2} of the DLCA 
procedure first proposed by Kolb and Hermann~\cite{kolb}.
The DLCA process is an iterative method which starts with a collection
of N identical spherical particles of diameter $\sigma_a$ randomly
placed in a cubic box of side L, with volume fraction 
$\eta=\frac{\pi}{6}\sigma_a^3\frac{N}{L^3}$. Aggregation proceeds via a 
diffusion motion of the particles.

If during their motion two clusters collide 
they stick together forming a new single aggregate.
The process is terminated when a single cluster 
is obtained.
Periodic boundary conditions are used at the edges of the simulation box.
Details of the DLCA algorithm can be found in references~\cite{kolb,hasmy2}.
We have generated with this algorithm configurations for
aerogels containing 515 particles in a box of 
edge $L=15 \sigma_a$, corresponding
to a volume fraction $\eta = 0.08$ and a 
porosity P, defined as the ratio between the free space volume and the volume
occupied by the gel, of 92\%.

\begin{figure}
\centering\epsfig{file=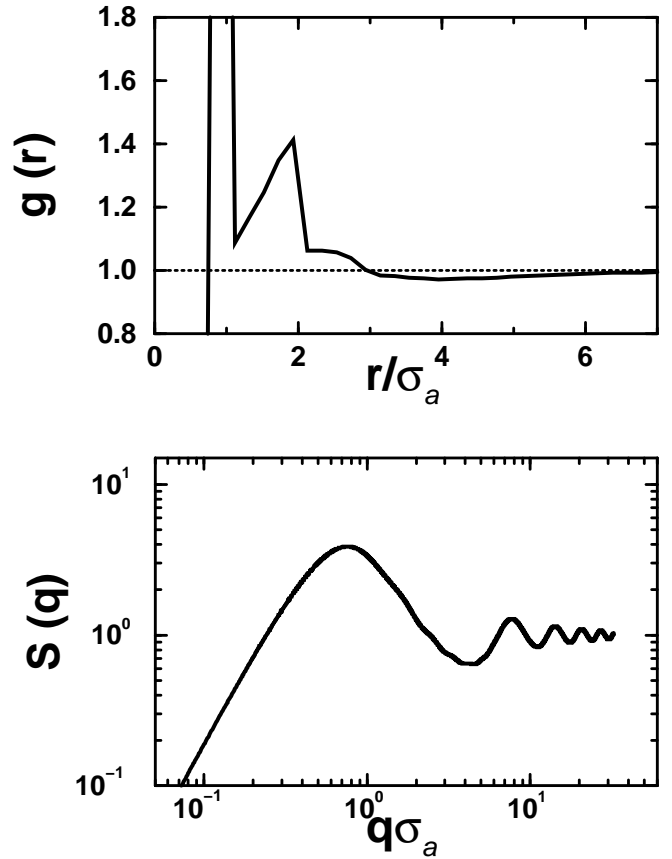,width=1\linewidth}  
\caption{
On the top plot
of $g(r)$ versus $r/\sigma_a$ for an aerogel containing 515 
particles in a box of edge $L=15 \sigma_a$.
This curve results from an average of 50 simulations.
On the bottom Log-Log plot of $S(q)$ versus $q\sigma_a$.}
\protect\label{fig:aero}
\end{figure}

We have calculated for these aggregates the radial distribution function g(r) 
and the static structure factor S(q) following the lines
of the simulation work by Hasmy and al.~\cite{hasmy2}
Our results are reported in Fig.~\ref{fig:aero}.
The radial distribution function g(r) presents a strong peak at 
$r=\sigma_a$, associated with
bonds between contacting particles, and other features
corresponding to short range correlations between particles
belonging to the same cluster 
(see references ~\cite{hasmy1,hasmy2} for further details).
Before reaching the asymptotic limit of 1, g(r) exhibits a
minimum corresponding to distances between particles located at the boundary 
of the clusters.
The location of this minimum gives an estimate of the mean clusters 
size $\xi$. In our case the mean cluster size is about $4 \sigma_a$.  
As a consequence of the presence of the minimum in g(r),
the scattering function S(q), which is related
to the Fourier transform of $g(r)-1$, exhibits a pronounced peak at small q.
The gel exhibits a fractal structure in 
the intermediate q-range. The
fractal dimension can be estimated from the 
power law behavior of the scattering function, $S(q) \sim q^{-D}$ in this 
q-range. In our case we have obtained $D\sim1.74$ consistent with the value
expected for silica aerogels in three dimensions~\cite{hasmy1,hasmy2}.

\section{SIMULATION DETAILS}

In our simulations 
the interactions between fluid particles are described
by a Lennard-Jones potential with a 
collision diameter $\sigma$ and a depth $\epsilon$ 
truncated at $r_c = 2.5\sigma$.
The aerogel particles are quenched and
interact with the fluid particles by means
of a purely repulsive hard sphere potential.
We assume that fluid particles have a collision diameter 
of the same size as the aerogel
hard core $\sigma=\sigma_a$.
The edge of our simulation box is $L=15\sigma$.
In the following Lennard-Jones units will be used: $\sigma$
for lengths,$\epsilon$ for energies and $\epsilon/k_B$ for temperatures.

In order to study the phase diagram of the 
Lennard-Jones fluid 
we have carried out Monte Carlo simulations in the grand canonical 
ensemble ~\cite{allen,frenkel} employing the
algorithm used by Wilding~\cite{wilding1,wilding2}.
MC moves consist of either an insertion or a deletion attempt, proposed 
with equal probability. Particles movements are implicitly implemented 
through insertion and deletion moves.

In order to calculate the liquid and vapor coexistence densities 
the chemical potential can be varied, at fixed temperature, 
until a bimodal shape of the particles number distribution P(N) 
is obtained.
The coexistence densities at the selected temperature 
correspond to the two peaks positions.
To obtain also the location of the liquid-gas saturation line 
in the ($\mu$,$T$) plane the
equal peak weight criterion~\cite{wilding2} can be applied. 
According to this criterion
the chemical potential is tuned, at constant temperature, 
until the measured P(N) is double-peaked with equal area under 
the two peaks. The corresponding chemical potential belongs 
to the fluid coexistence curve.

This conventional grand canonical ensemble Monte Carlo technique becomes
however impractical in the subcritical region we are interested in,
due to the large free energy  barrier 
separating the two phases and hindering spontaneous fluctuations of the system 
from the liquid to the gas phase and viceversa.
In order to enhance crossing of this free energy barrier we have used 
the recently proposed MES~\cite{berg,wilding1,wilding2}. 
With this powerful technique sampling
is made from a non Boltzmann distribution, 
with a modified Hamiltonian $H^{'} = H + g(N)$, where 
$ H ({\textbf{r}},N) = E({\textbf{r}}) - \mu N$ is the 
configurational Hamiltonian.
The biased sampling function $g(N)$ has to be chosen in such a way that 
the measured distribution $P^{'}(N)$ is nearly flat.
In this way the mixed-phase configurations will be sampled with 
approximately the same probability as the gas and liquid configurations.
The best choice for the biased sampling function
would be $g(N)=lnP(N)$, where
$P(N)$ is the distribution we are looking for. 
An estimate $\tilde{P}(N)$ can be obtained by extrapolation making use of the 
histogram reweighting technique~\cite{ferrenberg}.
In this approach we can accumulate the joint probability 
distribution of system energy and
particles number for a thermodynamical state characterized 
by a temperature $T_0$
and a chemical potential $\mu_0$ 
(near the critical point where the interfacial 
tension is low)~\cite{alvarez}:
\begin{equation}
\displaystyle P(N,E|T_0,\mu_0)= \frac{e^{-\frac{H_0}{kT_0}}
{\cal{D}}}{{\cal{Z}}_0}
\end{equation}
where ${\cal{D}}(N,E)$ is the density of
states and ${{\cal{Z}}_0}(T_0,\mu_0)$ is the grand partition function.
An estimate $\tilde{P}$ of this distribution
for another thermodynamical state $(T_1,\mu_1)$ can now be provided:
\begin{equation}
\displaystyle \tilde{P}(N,E|T_1,\mu_1)= \frac{{\cal{Z}}_0}{{\cal{Z}}_1} 
e^{-(\frac{H_1}{kT_1}-\frac{H_0}{kT_0})} P(N,E|T_0,\mu_0)
\end{equation}

The new thermodynamical point $(T_1,\mu_1)$ has
to be sufficiently close to $(T_0,\mu_0)$ in such a way that the
statistical weight of the new configuration is not
too different from the previous one.
Integrating over the system energy E, we can finally 
obtain a suitable bias function :

\begin{equation}
\displaystyle \tilde{P}(N|T_1,\mu_1)= \int dE \tilde{P}(N,E|T_1,\mu_1)
\end{equation}

At the end of the Monte Carlo simulation at $(T_1,\mu_1)$ with the biased
Hamiltonian we obtain the almost flat $P^{'}(N)$.
The real particles number distribution 
$P(N)$ pertaining to $(T_1,\mu_1)$ is recovered from:
$P(N)=P^{'}(N) \cdot \tilde{P}(N)$.

This method has given more accurate results than
the well known Gibbs ensemble technique and 
Gibbs-Duhem integration scheme both for bulk 
and confined fluids~\cite{wilding1,alvarez}.

We have performed extensive MES simulations in order to 
calculate the confined fluid phase diagram for a range of 
subcritical temperatures from $T=0.96$ to $T=0.79$.
Our simulations involved about $200 \cdot 10^6$ 
steps for system equilibration 
and, depending on temperature, from $1 \cdot 10^9$ 
to $2 \cdot 10^9$  steps for calculating 
the ensemble averages. 

Runs of this order of magnitude are necessary in order
to supply enough indipendent samples from each of the two phases
and consequently obtain density
distribution functions with enough accuracy for
our analysis~\cite{wilding2}. We note that on decreasing temperature
the rate of the interchanges between the liquid and gas
phases becomes more rare and longer runs are needed.

\section{RESULTS}

\subsection{Simulation in the two-phase region}

\begin{figure}[ht]
\centering\epsfig{file=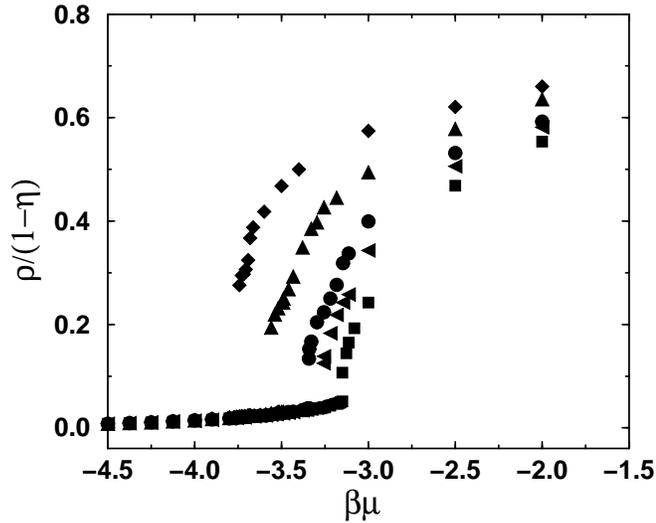,width=1\linewidth}  
\caption{Adsorption isotherms for a Lennard-Jones 
fluid in a DLCA aerogel for
a range of subcritical temperatures ($T=0.80$ diamonds, $T=0.85$ 
triangles up, $T=0.90$ circles, $T=0.92$ 
triangles left and $T=0.95$ squares).
The fluid density has been normalized by the void 
fraction $(1-\eta)$ of the
simulation box volume, where $\eta$ is the gel volume fraction.
All quantities are in Lennard-Jones units.}
\protect\label{fig:isot}
\end{figure}

First of all
we need to locate the coexistence curve of the confined fluid 
and in particular the near-critical region. For this sake 
we have calculated with standard grand canonical simulations
adsorption isotherms of the density $\rho$ versus the 
chemical potential $\mu$ for temperatures ranging 
from $T=0.80$ to $T=0.95$.
Our results are shown in Fig.~\ref{fig:isot}.
FIGURA2
For a given temperature, we observe
evidence of a coexistence at the same chemical
potential between a dilute 
gas and a medium density liquid in our confined system.
We have not observed in our isotherms the occurrence of a liquid-liquid
phase coexistence, at variance with 
previous computer simulation studies on the 
phase behavior of Lennard-Jones fluids in disordered 
non fractal matrices of spheres~\cite{monson1,alvarez}.

In our calculations we have not considered desorption
isotherms and the
hysteresis associated usually with capillary
condensation~\cite{gelb,kierlik} since we are
interested in the coexistence between equilibrium phases.

We note that due to the slope of the liquid portion of the
isotherms a small change of the chemical potential induces 
a strong variation in the liquid
density. Therefore
it is difficult to tune the chemical
potential with the appropriate resolution to locate
the coexisting densities with enough accuracy.
On approaching the critical region this problem becomes 
even worst.

\subsection{Multicanonical ensemble sampling}

In order to better investigate the coexistence region and
in particular the possible 
existence of a liquid-liquid phase coexistence we 
carried out extensive MES calculations.

\begin{figure}[ht]
\centering\epsfig{file=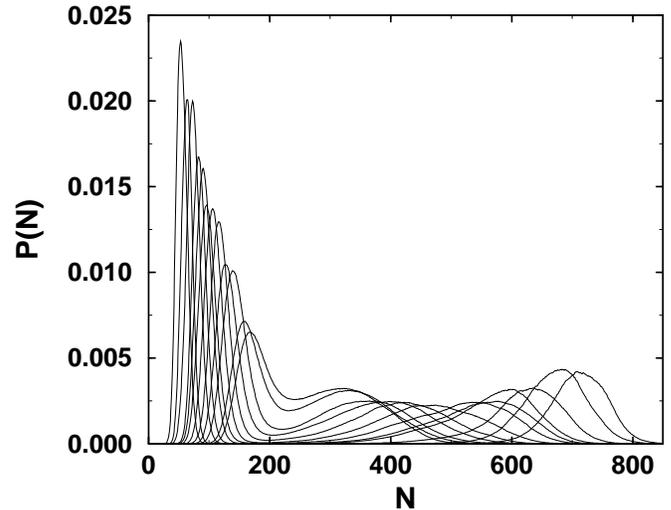,width=1\linewidth}    
\caption{Coexistence particles number distributions for subcritical
temperatures ranging from $T=0.96$ to $T=0.79$ in Lennard-Jones units. 
Curves with
closer peaks correspond to higher temperatures.}
\protect\label{fig:istog}
\end{figure}

From our isotherms calculations 
we have roughly estimated the region
where the onset of the bimodal shape of the particles number 
distribution P(N) can be observed. 
In our analysis we started from the double-peaked P(N)
obtained at the temperature $T=0.96$.
The histogram reweighting technique 
allowed us to estimate the coexistence liquid 
and vapor densities and chemical potential at this temperature.
Then by making use of the MES procedure 
we investigated a wide range of subcritical temperatures.   
In  Fig.~\ref{fig:istog} we report the distributions P(N) 
for the thermodynamical states investigated starting from
$T=0.96$.
 
At higher temperatures the 
liquid and gas have close densities and the 
double-peaked $P(N)$ distribution shows two partially overlapped peaks. 
As the temperature
decreases the difference between the coexisting vapor and liquid densities 
becomes more marked.

\begin{figure}[ht]
\centering\epsfig{file=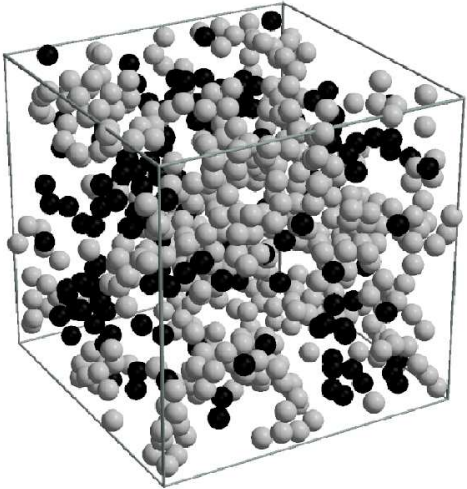,width=1\linewidth}    
\caption{A snapshot for the gas phase near the coexistence
at $T = 0.915$. 
The light grey and the black spheres
represent the gel and fluid particles, respectively. 
The boxlength is $L=15$.
All quantities are in Lennard-Jones units.}
\protect\label{fig:gas}
\end{figure}

\begin{figure}[ht]
\centering\epsfig{file=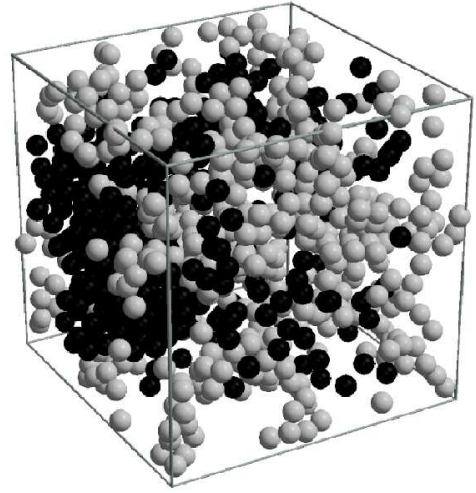,width=1\linewidth}        
\caption{A snapshot for the liquid 
phase near the coexistence at $T = 0.915$. 
The light grey and the black spheres
represent the gel and fluid particles, respectively.
The boxlength is $L=15$.
All quantities are in Lennard-Jones units.}
\protect\label{fig:liq}
\end{figure}

In the intermediate temperature
regime we do not observe the occurrence of a third peak associated
with a second coexisting liquid phase. We will comment more extensively 
about this point in the following.
Table~\ref{tab:1} summarizes the temperatures at which we performed our
simulations together with the values of the coexistence densities.
The MES algorithm allowed us to follow the
system down to $T=0.79$, where the liquid density is 
more than 10 times higher than the vapor one. 
\begin{table}
\begin{tabular}{|c|c|c|c|}
\hline
$T$ & ${\rho_g} / (1 - \eta)$ & ${\rho_l} / (1 - \eta)$ \\
\hline
0.960 &  0.054  &   0.104   \\     
0.950 &  0.051  &   0.107   \\   
0.930 &  0.045  &   0.117   \\ 
0.915 &  0.041  &   0.128   \\ 
0.900 &  0.038  &   0.138   \\ 
0.885 &  0.033  &   0.153   \\ 
0.870 &  0.031  &   0.172   \\  
0.860 &  0.029  &   0.185   \\ 
0.850 &  0.027  &   0.194   \\ 
0.830 &  0.023  &   0.205   \\  
0.810 &  0.021  &   0.220   \\  
0.790 &  0.017  &   0.228   \\
\hline   
\end{tabular}
\caption{The peak densities corresponding to the number particles
distributions shown in Fig.~\ref{fig:istog}.
All quantities are in Lennard-Jones units.}
\label{tab:1}
\end{table}
Snapshots from our simulations for the coexistence liquid and 
vapor configurations 
at one temperature are shown in Fig.~\ref{fig:gas} and  
Fig.~\ref{fig:liq}.
In both the coexisting phases the spatial distribution of 
the fluid molecules is 
inhomogeneous and the pores of the aerogel are far from 
being uniformly filled by the fluid.

\begin{figure}
\centering\epsfig{file=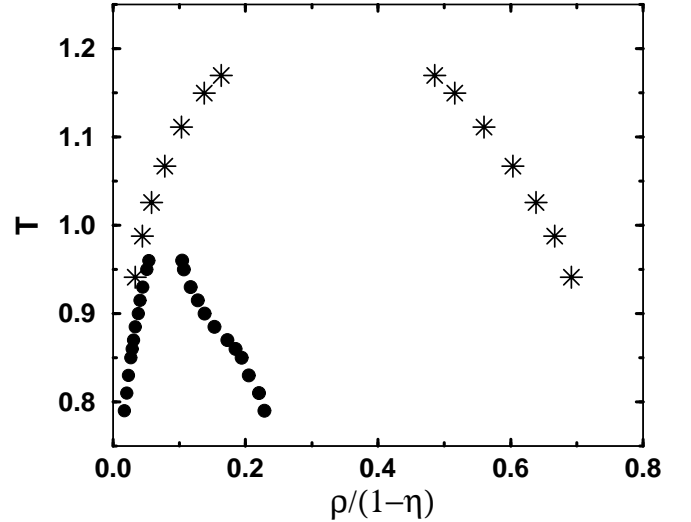,width=1\linewidth}        
\caption{Phase diagram for the confined fluid obtained from the particle
number distribution peak locations (circles) compared with that of
the bulk fluid~\cite{wilding1}(stars).
All quantities are in Lennard-Jones units.}
\protect\label{fig:coe}
\end{figure}

Fig.~\ref{fig:coe} shows the temperature versus density 
phase diagram we obtained
from the peaks locations in the P(N) distributions depicted 
in Fig.~\ref{fig:istog},
compared with that of the bulk fluid~\cite{wilding1}.
We find that the confined fluid phase diagram is substantially modified
by the presence of the aerogel : both the critical temperature and 
density are lower than in the bulk 
and the range of the vapor-liquid coexistence curve is much 
less extended. We recall that in the bulk the critical
parameters are: $T_c=1.1876$, $\rho_c=0.3197$ and 
$\mu_c=-2.778$~\cite{wilding1}.

Our findings about the gas-liquid coexistence properties 
are in qualitative agreement with previous 
computer simulation studies on Lennard-Jones 
fluids confined in random spheres matrices with purely repulsive
adsorbent-adsorbate interactions.~\cite{monson1,alvarez}.

None-the-less the phase diagram reported in Fig.~\ref{fig:coe} does not
present a liquid-liquid coexistence region.
It shows instead a well defined shoulder on the liquid side boundary 
for intermediate temperatures in the range investigated.

We must however stress that
confining primary particles in real aerogels are 
substantially larger than the fluid molecules,
therefore a study on a much larger system
could help to shed light on this issue.

\begin{figure}
\centering\epsfig{file=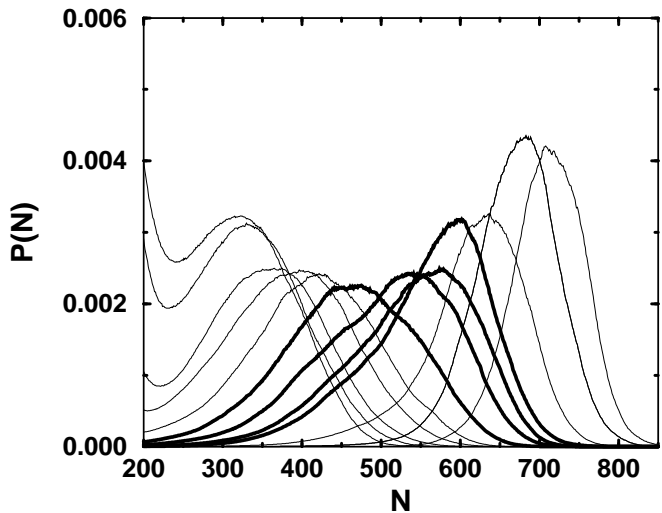,width=1\linewidth}        
\caption{Blow up of the 
liquid peaks of the number particles distributions depicted in 
Fig.~\ref{fig:istog}. The bold solid lines correspond to 
the liquid peaks generating the 
observed shoulder in the phase diagram ($T=0.885$, $T=0.87$, $T=0.86$ 
and $T=0.85$ in Lennard-Jones units).}
\protect\label{fig:liquid}
\end{figure}

In order to better define the nature of the liquid phase
for the thermodynamic points corresponding to this shouldering, 
in Fig~\ref{fig:liquid} we show a blow up of
the liquid peaks of the P(N) distributions 
centered around the intermediate temperature region.
The liquid peaks generating the shoulder in the coexistence curve are 
indicated with bold solid lines.
The liquid peaks shape shows a peculiar behavior with varying temperature.
For temperatures ranging from near the critical point to about $T=0.90$, the
shape of the liquid peaks is symmetric around the peaks maximum
as observed for the bulk~\cite{wilding1}.
For the reduced temperature corresponding to the onset of the shoulder,
$T=0.885$ (the first bold curve on the left),
the liquid peak shows a bump on
its right side that renders its shape slightly asymmetric.
We observe that the position of this bump roughly corresponds 
to the liquid peak maximum of the next $P(N)$
distribution, corresponding to $T=0.87$. 
For $T=0.87$ (the second bold curve
starting from the left) it is conversely observed a bump located
on the left side of the liquid peak at nearly the same position of the 
peak maximum measured at $T=0.885$.
We additionally note that the main peaks of these two curves 
show a more pronounced separation from each other 
compared with the other curves, in spite of being the temperature
jump between one curve and the next approximatively similar for
all the curves investigated.
The next two liquid peaks , corresponding to the end of the shouldering,
$T=0.86$ and $T=0.85$, are less asymmetric and closer than the two previous
ones.
For temperatures lower than $T=0.85$, after the shoulder in the phase diagram,
the shape of the liquids peaks returns to be 
symmetric and the liquid densities increase regularly on lowering 
temperature, similar to the high temperatures range.

From these observations we can infer that at high temperatures
the gas phase coexists with a medium density liquid.
Correspondingly the coexistence curve decreases regularly with temperature.
In the intermediate temperature range the 
liquid phase starts showing high density fluctuations and 
appears to coexists with a slightly more dense liquid.
This region is characterized by an asymmetry in the liquid peaks
of the $P(N)$ and correspondingly a shouldering of the
phase diagram, that mark a crossover to a slightly more dense phase.
For lower temperatures the higher density liquid becomes
the thermodynamically favoured phase coexisting with the vapor
and the coexistence curve resumes a regular descent.

Finally we report the liquid-vapor 
coexistence curve in the ($\mu$,$T$) 
plane as obtained from our multicanonical simulations.
According to the equal peak weight criterion, described in Sec.~3, 
the location of the liquid-vapor saturation 
line in this plane can be obtained looking for the values of temperature
and chemical potential at which the bimodal distributions $P(N)$ have 
equal area under the two peaks.
\begin{figure}
\centering\epsfig{file=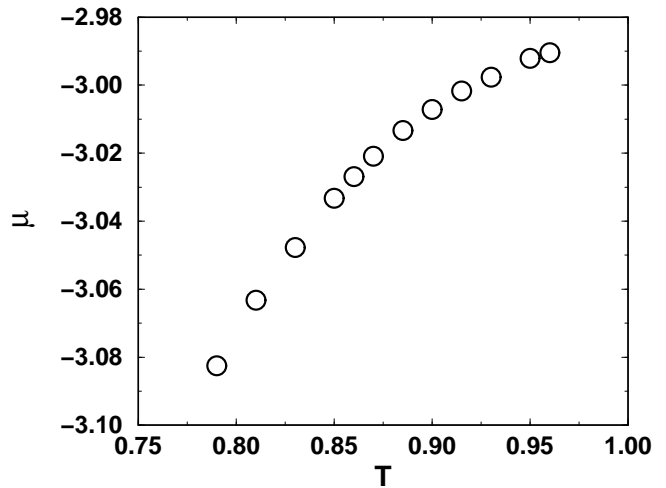,width=1\linewidth}        
\caption{The confined fluid liquid-gas saturation line in the 
($\mu$,$T$) plane for temperatures ranging 
from $T=0.96$ to $T=0.79$.
All quantities are in Lennard-Jones units.}
\protect\label{fig:mt}
\end{figure}
 
Fig.~\ref{fig:mt} shows the temperature behavior of the coexistence chemical 
potentials at which we obtained the P(N) distributions reported in 
Fig.~\ref{fig:istog}. Analogous to the bulk we observe a monotonic
regular behavior. Similar to the temperature
versus density phase diagram, there is a substantial shrinkage of 
the range of chemical potentials at which 
we observe the coexistence of the liquid and gas phases 
with respect to the bulk. 
In fact in the bulk~\cite{wilding1} for a similar interval of temperatures
as ours, i.e. $\Delta T = 0.17$, 
a range $\Delta \mu \sim 0.6$ is found, while 
in our confined system $\Delta \mu \sim 0.09$, more than six
times lower.

\section{CONCLUSIONS}

We have performed a computer simulation study of the 
phase behavior of a Lennard-Jones
fluid adsorbed in a highly porous fractal aerogel.
The gel environment has been generated with the DLCA 
algorithm~\cite{kolb,hasmy2} in order to obtain a more 
realistic confining structure than those reported in literature.
Besides the confined fluid phase diagram has been calculated
by performing  multicanonical
ensemble simulations in the framework of the 
Monte Carlo grand canonical ensemble 
technique~\cite{berg,wilding1,wilding2}. 

The proper bias has been found by applying the single-histogram
reweighting technique~\cite{ferrenberg}.
This procedure is particularly 
suitable for analysing the transition state of the first order
phase transitions in great detail.

We found that the phase diagram of the confined fluid  
is substantially modified with respect to the bulk by the presence 
of the gel: the critical temperature and density are lower and the liquid-gas
coexistence curve is much narrower than in the bulk.
Our findings about the vapor-liquid coexistence properties are in 
qualitative agreement with experimental observations
and previous theoretical studies.
We have not found a clear evidence for a second liquid-liquid coexistence 
region in the fluid phase diagram. 
The presence of a liquid-liquid phase coexistence
has been sometimes reported, depending on the particular configuration
generated for the adsorbent, random matrices of
spheres~\cite{monson1,alvarez}. 
Our coexistence
curve shows a shoulder on the liquid side boundary
in the intermediate temperature range.
For the same temperatures the liquid peaks shape of our 
particles number distributions are highly asymmetric and show a bump. 
However, since we do not find two distinctly resolved
peaks, we must conclude that in our case coexistence 
between two liquid phases is never observed.
We do observe instead a crossing of the liquid from a 
lower to an higher density phase upon decreasing temperature.

We have employed in this study a very sophisticated technique for
an accurate location of the coexistence curve.
Due to the fractal structure of the confining environment
the results should in principle not depend on the realization.
However we cannot exclude that finite size effects could
be present. Only extensive analysis with increasing sizes  
can confirm how sensitive the results are to the matrix
realization. 

In order to gain deep insight about 
the modification of the phase diagram of fluids confined
in aerogels the modeling of the adsorbent environment 
can be further improved
to make contact with the experimental situations.
It would be interesting to explore in  
a more carefully designed  confining structure with 
a network of multishaped voids if the 
two liquid phases will become distinguishable.
In particular the liquid-liquid coexistence is
smeared out in our
present simulation, where we observe only
a shouldering in the phase diagram.
It would be valuable to perform simulations 
on even larger systems to better investigate the role of
the fractal behavior of the confining medium and perform
an extensive analysis of finite size effects.

Work on these improvements to our present study is 
currently in progress.

\end{document}